# Efficient Network Traffic Feature Sets for IoT Intrusion Detection

Miguel Silva[0009-0008-6630-9939], João Vitorino[0000-0002-4968-3653], Eva Maia[0000-0002-8075-531X] and Isabel Praça[0000-0002-2519-9859]

Research Group on Intelligent Engineering and Computing for Advanced Innovation and
Development (GECAD), School of Engineering, Polytechnic of Porto (ISEP/IPP),
4249-015 Porto, Portugal
`{mdgsa,jpmvo,egm,icp}@isep.ipp.pt`

**Abstract.** The use of Machine Learning (ML) models in cybersecurity solutions requires high-quality data that is stripped of redundant, missing, and noisy information. By selecting the most relevant features, data integrity and model efficiency can be significantly improved. This work evaluates the feature sets provided by a combination of different feature selection methods, namely Information Gain, Chi-Squared Test, Recursive Feature Elimination, Mean Absolute Deviation, and Dispersion Ratio, in multiple IoT network datasets. The influence of the smaller feature sets on both the classification performance and the training time of ML models is compared, with the aim of increasing the computational efficiency of IoT intrusion detection. Overall, the most impactful features of each dataset were identified, and the ML models obtained higher computational efficiency while preserving a good generalization, showing little to no difference between the sets.

## 1 Introduction

The rapid expansion of the Internet of Things (IoT) is largely owed to the advances of communication technology and the increasing availability of devices. Due to their computational restraints, it is essential to improve the security and efficiency of hardware and networks within the IoT systems [1], [2]. Network Intrusion Detection Systems (NIDS) are tools used to monitor network traffic, identifying malicious behavior and prevent potential attacks in order to maintain the confidentiality, integrity and availability of the network traffic [3]. These systems are able to analyze network activity with the help of Artificial Intelligence (AI), namely Machine Learning (ML) models, which classify network traffic as either benign or malicious [4].

To train reliable and robust ML models, it is essential to have accurate information that truthfully represents the network activity [5]. Due to IoT's large amount of data and the limited resources available to process it, the inclusion of less impactful features can degrade its robustness and lower the detection speed [6], raising energy concerns. Therefore, improving the classification efficiency of ML models requires a careful balance between maximizing their generalization and minimizing the utilization of less impactful data [7] - [9].

This study aims to increase the computational efficiency of the ML models used in IoT NIDS, by reducing the dimensionality of the utilized datasets. Five feature selection techniques, Information Gain, Chi-squared Test, Recursive Feature Elimination, Mean Absolute Deviation, and Dispersion Ratio, were used to obtain the most relevant features of the Bot-IoT, IoT-23 and Ton-IoT datasets. The classification performance and the training time of multiple ML models was evaluated and compared. Three different ML models, Random Forest (RF), Extreme Gradient Boosting (XGB), and Light Gradient Boosting Machine (LGBM), were trained using both the full feature set and the subset with only the most relevant features of each dataset.

The present paper is structured in following sections: Section 2 provides a survey of previous work using feature selection in intrusion detection. Section 3 describes the considered datasets, feature selection methods, and ML models. Section 4 presents and discusses the obtained results. Section 5 addresses the main conclusions and future research topics.

## 2    Related Work

To perform an impactful feature selection process, it is essential to have a comprehensive understanding of the conclusions of previous work. The number of features and predictors of an ML model should be minimized, both as a dimensionality reduction approach to improve their computational efficiency [10] and as a feature squeezing defense to improve their robustness [11]. Since the datasets used in network intrusion detection are commonly very large, a smaller feature set can be significantly beneficial to reduce time consumption and financial cost of training and testing ML models [12].

S. Krishnaveni et al. [13] introduced a feature selection methodology using five filtering feature selection techniques, namely Information Gain, Chi-Squared Test, Gain Ratio, Symmetric Uncertainty, and Relief. The study used the Honeypot real-time dataset, the NSL-KDD dataset, and the Kyoto dataset, ranking the features and discarding those that scored less than 20%. Multiple models were trained and evaluated, resulting in a majority voting ensemble with an accuracy of 96.06%.

In another study [14], the Chi-Squared Test, Mutual Information Statistic, and a correlation analysis were used to assess the importance and relevance of the features found in the KDD Cup 99 dataset. This analysis resulted in 11 features being discarded, that were found to be redundant. The selected features were used to train a Deep Neural Network (DNN) model, which achieved 99.4% accuracy and 98.8% F1-score, surpassing the performance of the same model with the whole dataset.

Another study [15] aimed at reducing the dimension and removing redundancy to improve the stability and accuracy of intrusion detection systems with low computational and time complexity, using a hybrid method combining Correlation-based Feature Selection (CFS) and the Bat Algorithm (BA). This method was applied to the NSL-KDD dataset, the AWID dataset, and the CIC-IDS2017 dataset, resulting in sets of 10, 8, and 13 features, respectively. By using ensemble classifiers with these feature sets, the models achieved accuracies of 99.81%, 99.52% and 99.89%.

In summary, previous works successfully used feature selection to optimize IoT datasets without addressing training efficiency. This study implements different techniques and evaluates both ML and training time efficiency.

## 3 Methodology

This section describes the considered datasets, selection methods, and models. The study was carried out on a machine with 16GB of RAM, a 6-core CPU and an 8GB GPU. The implementation was done with the Python programming language and the following libraries: *numpy* and *pandas* for general data manipulation, *xgboost* for the implementation of XGB, *lightgbm* for LGBM, and *scikit-learn* for the implementation of RF and the feature selection methods.

### 3.1 Datasets and Selection Methods

The datasets included in this study contain labeled network traffic flows that represent normal traffic and malware attacks targeting IoT networks. These flows are classified as either benign, sent as part of normal network operations, or malicious, send by an attacker with malicious intent.

The IoT-23 dataset [16] was created by the Stratosphere Research Laboratory and contains 23 captures of the network activity of a real IoT network, using smart devices with access to an internet connection. On the other hand, the Bot-IoT dataset [17] contains network traffic generated from a realistic IoT network environment simulation developed at the Intelligent Security Group of the University of New South Wales, with simulated devices being infected by malware and controlled by botnets. In turn, the more recent Ton-IoT dataset [18] was created from an improved network environment simulation at the same University. Even though only network traffic was used in this study, the dataset also includes Windows and Linux audit traces, which could be valuable for future integration of network-based with host-based detection [19].

For each of the three datasets, five feature selection methods were applied independently, and then their results were combined to obtain a single ranking of the most relevant features of each dataset. The considered methods were: **Information Gain**, which is broadly used to analyze the change in entropy when a known class is removed, providing information on the relevance of a feature [20]; **Chi-Squared Test**, measuring the degree of dependence between a term and a class, fitted to the chi-squared distribution with one degree of freedom for analysis [21]; **Recursive Feature Elimination**, which reduces the number of features based on the weights assigned by a classification model [22]; **Mean Absolute Deviation**, providing a scale factor in the Laplace distribution, serving as a measurement of the dispersion inherent in a feature, which is often used as a substitute for the standard deviation [23]; **Dispersion Ratio**, which is defined as the square root of the ratio between two elements, where the numerator represents the variance between different categories, while the denominator represents the variance of that feature in the dataset [24].

### 3.2 Models and Fine-tuning

Three types of ML models were selected: RF, XGB and LGBM [25]. The optimal configuration for each model in each dataset was obtained through a grid search involving well-established hyperparameter combinations [26]. The best configurations were selected through a 5-fold cross-validation, with evaluation based on the macro-average F1-score to address the imbalance observed in the datasets. The models and their fine-tuned hyperparameters are described below.

**Random Forest**. RF [27] is a supervised ensemble of decision trees, where the predictions of each tree are combined to choose a class through the wisdom of the crowd. Table 1 summarizes the configuration.

Table 1. Summary of RF configuration.

| Parameter | Value |
|---|---|
| Criterion | Gini Impurity |
| No. of estimators | 100 |
| Max. features | $\sqrt{No.\ of\ features}$ |
| Max. depth of a tree | 16 |
| Min. samples in a leaf | 2 |

**Extreme Gradient Boosting**. XGB [28] performs gradient boosting using a supervised decision tree ensemble. The Histogram method is used to choose the best node splits. The configuration is summarized in Table 2.

Table 2. Summary of XGB configuration.

| Parameter | Value |
|---|---|
| Loss function | Cross-Entropy |
| Learning rate | 0.2 |
| No. of estimators | 80 to 100 |
| Feature subsample | 0.7 to 0.8 |
| Min. loss reduction | 0.01 |
| Max. depth of a tree | 8 |

**Light Gradient Boosting Machine**. LGBM [29] also performs gradient boosting with a supervised tree ensemble. The trees are constructed using Gradient-based One-Side Sampling (GOSS). The configuration is presented in Table 3.

Table 3. Summary of LGBM configuration.

| Parameter | Value |
|---|---|
| Loss function | Cross-Entropy |
| Learning rate | 0.01 to 0.2 |
| No. of estimators | 100 to 120 |
| Feature subsample | 0.7 |
| Min. loss reduction | 0.01 |
| Max. leaves in a tree | 32 |
| Min. samples in a leaf | 16 |

## 4 Results and Discussion

This section presents and discusses the results obtained from the feature selection and the evaluation of the ML models, highlighting the most relevant features of each dataset. The importance values obtained from each feature selection method were normalized to obtain a single value in the range of zero to one hundred percent, and the 8 most relevant features were chosen.

The evaluation metrics used to compare the ML models were accuracy (ACC), precision (PRC), recall (RCL), F1-score (F1S), and false positive rate (FPR). Since the accuracy is affected by class imbalance between benign and malicious flows, the F1-score metric provides the most comprehensive view of the generalization of an ML model. The optimal results would be 100% for all metrics except FPR, which should be as close to 0% as possible [25].

### 4.1 Bot-IoT

Regarding the Bot-IoT dataset, the most relevant features collectively represent 92% of the total importance. The importance of the three most relevant features exceeds half this value, highlighting their overall importance on the set. Features related to bytes got a significant consideration, accounting for a total of 42%. The total packets per second was found to be the most relevant feature, whereas the packets sent from the client to the server had less than 2% of importance. Table 4 provides the value of each selected feature of Bot-IoT, with Fwd corresponding to Forward and Bwd to Backward.

Table 4. Feature ranking for Bot-IoT.

| Value (%) | Feature | Value (%) | Feature |
|---|---|---|---|
| 21.59 | Packets Per Second | 11.50 | Bwd. Bytes |
| 18.06 | Total Bytes | 7.43 | Communication Protocol |
| 12.70 | Flags | 6.89 | Destination Port |
| 12.40 | Fwd. Bytes | 1.28 | Fwd. Packets per second |

For all evaluation metrics, the models trained with the Bot-IoT dataset achieved very good performances. The feature selection process significantly improved the training time for all three models, although with some minor differences compared to using the entire dataset. With the selected features, the RF model was able to improve its F1S, however, the LGBM model suffered a drop in scoring. The RF model performed best and showed the largest difference in training time, meanwhile the LGBM model suffered a 3% increase in FPR using only the selected features, which could lead to false alarms in real computer network scenarios. The XGB model maintained almost the same overall score using only the selected features, but slightly increased its FPR. It is worth noting that in this context, 0.001% corresponds to a total of 2 flows. Table 5 provides the results of the models trained with and without the selected features.

Table 5. Obtained results for Bot-IoT.

| Model | Feature Selection | Evaluation Metrics (%) | | | | | Training Time |
|---|---|---|---|---|---|---|---|
| | | ACC | PRC | RCL | F1S | FPR | |
| RF | No | 99.993 | 99.996 | 99.998 | 99.996 | 6.2937 | 14.581 |
| | Yes | 99.994 | 99.996 | 99.998 | 99.997 | 6.2937 | 10.873 |
| XGB | No | 99.988 | 99.996 | 99.992 | 99.994 | 6.2937 | 5.0667 |
| | Yes | 99.989 | 99.995 | 99.994 | 99.994 | 6.9930 | 3.0130 |
| LGBM | No | 99.990 | 99.994 | 99.995 | 99.995 | 7.6923 | 3.1380 |
| | Yes | 99.987 | 99.992 | 99.994 | 99.993 | 10.490 | 2.6019 |

### 4.2   IoT-23

Regarding the specific feature set for IoT-23 dataset, the selected features accounted for more than 96% of the total relevancy. Two features stood out as notably decisive, adding up to more than half of the total relevance, with a combined score of more than 55%. These features cover the total number of bytes transmitted from the client machine to the server, with and without the header. In comparison, the importance scores the bytes of the packets transmitted from the server back to the client, with and without the header, is significantly lower, approximately 8 times less. This gap underlines the importance attached to packets sent from the client machine to the server in the dataset. Table 6 provides the value of each selected feature of IoT-23.

Table 6. Feature ranking for IoT-23.

| Value (%) | Feature | Value (%) | Feature |
|---|---|---|---|
| 32.54 | Fwd. Bytes | 8.45 | Communication Protocol |
| 22.99 | Fwd. Bytes w/ Header | 4.98 | Bwd. Bytes w/ Header |
| 11.60 | Destination Port | 1.77 | Bwd. Bytes |
| 12.38 | Flags | 1.58 | Application protocol |

The results of all the models trained with the IoT-23 dataset achieved very good performances, with minor differences across all the metrics. The time required to train each of the models decreased using only the top 8 features, although the results suffered from 0 to 0.0001% changes. The LGBM model achieved the best overall score, not only having the same F1R as the other best performing model in this metric, the RF, but also achieving better ACC, PRC, RCL and FPR, while requiring the least time to train, although the change in time was not as large as for the RF. The XGB model almost maintained its overall score, while reducing its training time to almost two-thirds when using only the selected features. It is worth noting that in this context, 0.0001% corresponds to a total of 3 flows. Table 7 provides the results of the models trained with and without the selected features.

Table 7. Obtained results for IoT-23.

| Model | Feature Selection | Evaluation Metrics (%) | | | | | Training Time |
|---|---|---|---|---|---|---|---|
| | | ACC | PRC | RCL | F1S | FPR | |
| RF | No | 99.992 | 99.992 | 99.993 | 99.993 | 0.0091 | 17.231 |
| | Yes | 99.991 | 99.992 | 99.991 | 99.992 | 0.0091 | 15.002 |
| XGB | No | 99.991 | 99.992 | 99.992 | 99.992 | 0.0091 | 6.0737 |
| | Yes | 99.992 | 99.992 | 99.993 | 99.992 | 0.0091 | 4.2814 |
| LGBM | No | 99.993 | 99.993 | 99.993 | 99.993 | 0.0084 | 3.9694 |
| | Yes | 99.992 | 99.993 | 99.992 | 99.992 | 0.0084 | 3.5375 |

### 4.3 Ton-IoT

Regarding the Ton-IoT dataset, the set of features obtained through the feature selection methods showed an overall importance of 93%. The two features considered to be more relevant collectively account for more than 59% of the total importance score, representing more than half of the overall importance. These features are both related to the bytes in the flow, originating from the client machine to the server and the response of the server back to the client, with a difference of less than 7%. In contrast, the features related to the bytes transmitted that include the header have a much smaller combined score of more than 10%, which account for almost a sixth of the total value without the header. These considered results highlight the importance of the two top features, while underlining the low relevance of the header of the packets. Table 8 provides the value of each selected feature of Ton-IoT.

Table 8. Feature ranking for Ton-IoT.

| Value (%) | Feature | Value (%) | Feature |
|---|---|---|---|
| 32.97 | Fwd. Bytes | 5.23 | Destination Port |
| 26.60 | Bwd. Bytes | 4.87 | Missed Bytes |
| 9.02 | Flags | 4.50 | Bwd. Bytes w/ Header |
| 6.02 | Fwd. Bytes w/ Header | 3.99 | Communication Protocol |

Despite significantly longer training times, due to the size of the dataset, the models trained with the Ton-IoT dataset achieved good scores. For all three models, a slight decrease in the evaluation metrics was measured, which accompanied a significant reduction in training time, except for the LGBM model. Despite using fewer features, the LGBM required more training time and achieved the lowest score among the models. Despite a minor 0.001% decrease of F1S and a slight increase of FPR, the RF model significantly reduced training time. Since RF is the model with the longest training time in this context, this is particularly relevant. It is worth noting that in this context, 0.001% corresponds to 67 flows, which is considerably higher when compared to the other datasets. Table 9 provides the results of the models trained with and without the selected features.

Table 9. Obtained results for Ton-IoT.

| Model | Feature Selection | Evaluation Metrics (%) | | | | | Training Time |
|---|---|---|---|---|---|---|---|
| | | ACC | PRC | RCL | F1S | FPR | |
| RF | No | 99.920 | 99.935 | 99.982 | 99.958 | 1.7498 | 1149.0 |
| | Yes | 99.917 | 99.933 | 99.981 | 99.957 | 1.8001 | 957.06 |
| XGB | No | 99.908 | 99.943 | 99.962 | 99.952 | 1.5453 | 162.48 |
| | Yes | 99.892 | 99.928 | 99.960 | 99.944 | 1.9405 | 154.92 |
| LGBM | No | 99.877 | 99.919 | 99.953 | 99.936 | 2.1789 | 78.614 |
| | Yes | 99.829 | 99.889 | 99.933 | 99.911 | 2.9955 | 81.074 |

## 5  Conclusions

This study evaluated the feature sets provided by a combination of Information Gain, Chi-Squared Test, Recursive Feature Elimination, Mean Absolute Deviation, and Dispersion Ratio. These methods were applied to the Bot-IoT, IoT-23, and Ton-IoT datasets. Different ML models, RF, XGB, and LGBM, were trained and evaluated using all the available data and a smaller feature set containing only the 8 most relevant features.

The feature selection process identified the features that were decisive for distinguishing between benign and malicious network flows. For each dataset, the smaller feature set was compared with the full feature set, to evaluate the trade-off between improving computational efficiency by reducing the features and improving the overall model score by including all available features.

The three ML models performed well in all metrics, using both the selected features and all available data. Since the information of each dataset represents different IoT networks, the results show that the selected features allow a good transferability between datasets, as they achieved a score almost as high as using all available features. Even though LGBM is a more complex model than RF, it achieved similar results and a shorter training time for all datasets.

While the IoT-23 dataset achieved the best results across multiple evaluation metrics, it is worth noting that training models with selected features can occasionally result in higher false positive rates or longer training times. It is crucial to evaluate the performance of an ML model on different datasets prior to deployment. This not only guarantees a robust generalization, but also ensures that the model achieves a reasonable balance between accuracy and efficiency, leading to its successful application in real-world scenarios.

In future works, it could be valuable to explore different types of ML and deep learning models to evaluate their robustness and computational efficiency when using smaller feature sets, across multiple IoT network datasets.

**Acknowledgments.** This work was supported by the CYDERCO project, which has received funding from the European Cybersecurity Competence Centre under grant agreement 101128052. This work has also received funding from UIDB/00760/2020.